\documentclass[aps,prd,10pt,notitlepage,nofootinbib]{revtex4-1}

\usepackage[utf8]{inputenc}
\usepackage{amsmath,amssymb,amsfonts}
\usepackage{graphicx, overpic}
\usepackage{palatino}

\usepackage{bbold}
\usepackage{bm}
\usepackage[usenames,dvipsnames]{xcolor}
\usepackage{color}
\usepackage[colorlinks=true,linkcolor=blue,urlcolor=blue,citecolor=blue]{hyperref}

\usepackage{slashed}
\usepackage[english]{babel}
\usepackage{dcolumn}
\usepackage{pifont}
\usepackage{dsfont,mathrsfs}
\usepackage{cancel}
\usepackage{bigints}
\usepackage{accents}
\usepackage{soul}
\usepackage{multirow}
\usepackage{natbib}
\usepackage{tikz-feynman}

\usepackage{amsmath, amssymb}
\usepackage{bbold}
\usepackage{makecell}
\usepackage{simpler-wick}
\usepackage[export]{adjustbox}

\usepackage[bb=boondox]{mathalpha}

\usepackage{tikz}
\usepackage{tikz-feynman}
\tikzfeynmanset{compat=1.1.0} 

\usepackage{orcidlink} 

\DeclareMathOperator*{\SumInt}{%
	\mathchoice%
	{\ooalign{$\displaystyle\sum$\cr\hidewidth$\displaystyle\int$\hidewidth\cr}}
	{\ooalign{\raisebox{.14\height}{\scalebox{.7}{$\textstyle\sum$}}\cr\hidewidth$\textstyle\int$\hidewidth\cr}}
	{\ooalign{\raisebox{.2\height}{\scalebox{.6}{$\scriptstyle\sum$}}\cr$\scriptstyle\int$\cr}}
	{\ooalign{\raisebox{.2\height}{\scalebox{.6}{$\scriptstyle\sum$}}\cr$\scriptstyle\int$\cr}}
}

\newcommand{\nn}{\nonumber}

\newcommand{\MB}[1]{\left|#1\right|}
\newcommand{\FB}[1]{\left(#1\right)}
\newcommand{\SB}[1]{\left\{#1\right\}}
\newcommand{\TB}[1]{\left[#1\right]}

\newcommand{\IM}{\text{Im}}

\newcommand{\half}{\dfrac{1}{2}}
\newcommand{\gm}{\gamma}

\newcommand{\Tr}[1]{{\rm Tr}\TB{#1}}

\newcommand{\mS}{\mathcal{S}}

\newcommand{\threeint}[1]{\int \frac{d^3 {#1}}{(2\pi)^3}}

\newcommand{\Pp}{\mathds{P}_+}

\newcommand{\Pm}{\mathds{P}_-}

\newcommand{\FSumInt}[1]{\SumInt_{\SB{#1}}}

\newcommand{\retPi}{^{\rm ret}\Pi^\mu_\mu (P)}
\newcommand{\ovK}{\overline{K}}
\newcommand{\ovQ}{\overline{Q}}
\newcommand{\w}{\omega}
\renewcommand{\wp}{\omega^{\prime}}

\allowdisplaybreaks

\begin{document}
	\title{Soft-contribution to thermal photon emission from chiral QCD medium}

	
	\author{Nilanjan Chaudhuri\orcidlink{0000-0002-7776-3503}$^{a}$}
	\email{n.chaudhri@vecc.gov.in}
	\email{nilanjan.vecc@gmail.com}
	\author{Sourav Duari\orcidlink{0009-0006-0795-5186}$^{a,c}$}
	\email{s.duari@vecc.gov.in}
	\email{sduari.vecc@gmail.com}
	
	\author{Pradip Roy\orcidlink{0009-0002-7233-4408}$^{b,c}$}
	\email{pradipk.roy@saha.ac.in}	
	
	\author{Sourav Sarkar\orcidlink{0000-0002-2952-3767}$^{a,c}$}
	\email{sourav.vecc@gmail.com}

	\affiliation{$^a$Variable Energy Cyclotron Centre, 1/AF Bidhannagar, Kolkata - 700064, India}
	\affiliation{$^b$Saha Institute of Nuclear Physics, 1/AF Bidhannagar, Kolkata - 700064, India}
	\affiliation{$^c$Homi Bhabha National Institute, Training School Complex, Anushaktinagar, Mumbai - 400085, India}
%
	
	\begin{abstract}
			We evaluate the thermal photon emission rate from a chirally asymmetric quark gluon plasma using the Hard Thermal Loop approximation.
		The quasiparticle and plasmino modes prevalent at finite temperature split into $L$ and $R$-modes in the presence of chiral imbalance and are found to disperse differently acquiring different thermal masses. 
		 The soft contribution to the thermal photon emission rate obtained from the retarded self-energy is found to contain additional terms proportional to the square of the quark and chiral chemical potentials which is found to cause an enhancement to thermal photon emission in the presence of chiral imbalance.

	\end{abstract}
	
	\maketitle
		\section{Introduction}

In QCD the compactness of the non-Abelian $SU(3)$ gauge group allows for topologically nontrivial configurations of the gluon field. This leads to a vacuum structure comprising of the superposition of an infinite number of  topologically distinct states which are characterised by non-zero winding number~\cite{Shifman:1988zk,Lenz:2001me}. These distinct vacua are connected by tunneling instanton transitions through a potential barrier whose height is of the order of the QCD scale, $\Lambda_{\rm QCD}$~\cite{Belavin:1975fg,tHooft:1976rip,tHooft:1976snw}. Topological objects in QCD are likely to play an important role in many fundamental non-perturbative processes such as chiral symmetry breaking and possibly confinement. These objects are not only important for vacuum physics but are also relevant at high temperatures such as those realized in heavy-ion collisions at RHIC and LHC energies due to the substantial production of another class of objects known as sphalerons~\cite{Manton:1983nd,Klinkhamer:1984di}. It is conjectured that the high abundance of sphalerons enable thermally activated transitions over the energy barrier separating topologically distinct vacua~\cite{Kuzmin:1985mm,Arnold:1987mh,Khlebnikov:1988sr,Arnold:1987zg}. These topologically nontrivial gauge field configurations can induce $P$ and $CP$ odd effects. Through the axial anomaly of QCD such processes can lead to the imbalance between left- and right-handed quarks~\cite{Adler:1969gk,Bell:1969ts} which is related to the topological winding number  : $N_R - N_L
 = -2\, Q_w $~\cite{Fukushima:2008xe}. Although no global \(CP\) violation has been experimentally observed such local violations may lead to domains of finite chiral imbalance in the QGP~\cite{McLerran:1990de,Moore:2010jd}. This imbalance is commonly characterized by a chiral chemical potential \( \mu_5 \), which quantifies the difference in number densities between right- and left-handed quarks.

Recently, anomalous transport phenomena have gained significant attention  with implications spanning from high-energy nuclear physics to condensed matter systems~\cite{Vilenkin:1980fu,Kharzeev:2013ffa,Miransky:2015ava,Kharzeev:2022ydx}. Of particular importance is the Chiral Magnetic Effect (CME)~\cite{Kharzeev:2007jp,Fukushima:2008xe,Kharzeev:2012ph,Kharzeev:2015znc,Koch:2016pzl} which generates a vector current in the direction of the applied magnetic field. The CME offers a promising avenue to probe topological fluctuations in QCD~\cite{Kharzeev:2020jxw}. Similar mechanisms have also been explored in Dirac and Weyl semimetals~\cite{Son:2012bg,Gorbar:2013dha,Li:2014bha,Cortijo:2016wnf,Kaushik:2018tjj,Sukhachov:2021fkh} where chiral quasiparticles are realized in condensed matter systems.
 
 Electromagnetic probes such as photons and dileptons have been long considered to be the most effective probes to characterize the initial state of the matter produced in relativistic heavy ion collision experiments at RHIC and LHC. Owing to their long mean free path compared to the dimension of the system photons and dileptons escape the system without any interaction and therefore carry
 information about the conditions of the system from where they are produced.
 In Ref.~\cite{Chaudhuri:2022rwo} the dilepton production rate (DPR) has been calculated in a hot and dense chiral medium and it is shown that the DPR is
 enhanced in the low invariant mass region due to the contribution from the Landau cut. As a result the forbidden gap between the Unitary cut and the Landau cut vanishes at high temperature and chiral chemical potential. Interestingly, spin polarized photon and dilepton emission from a chiral plasma have been conjectured to provide evidence for possible $P$ and $CP$ violations in strong interaction~\cite{Mamo:2013jda}. Again, a novel mechanism for soft photon production based on conformal anomaly of QCD and QED along with the existence of strong electromagnetic fields in heavy collision has been proposed~\cite{Basar:2012bp}. This mechanism is also shown to provide a significant positive contribution to the azimuthal anisotropy of photons as well as to the radial flow.
 
 In this work we calculate the emission rate of thermal photons from a quark gluon plasma with local chiral imbalance. In particular, we focus on the soft contribution to the production rate using the hard thermal loop (HTL) resummed fermion propagator. To this end, we calculate the one loop retarded photon self energy with one effective propagator and one bare propagator in the loop. The imaginary part of the trace of the retarded self energy is related to the photon production rate. A separation scale $k_c$, where $gT \ll k_c \ll T$, has been introduced in order to render the integration over loop momenta finite. Hence it is expected that the results will be scale dependent. However, it should be noted that in order to be phenomenologically relevant, in addition to the soft contribution the photon production rate requires contribution from hard scatterings coming from quark and anti-quark annihilation and (QCD) Compton processes. At zero density, using Boltzmann approximation it is possible to show analytically that on adding the hard and soft contributions the separation scale $k_c$ does not appear in the final expression~\cite{Kapusta:1991qp,Baier:1991em}. At finite chemical potential Bose and Fermi distributions have to be used to match the soft and hard contributions in which case the hard contribution can only be determined numerically and the final result becomes independent of $k_c$ as shown in~\cite{Traxler:1994hy}. However, as argued in Ref.~\cite{Hou:1996uq} most of the contribution comes from the soft part in the weak coupling limit. Thus we will ignore the hard contribution in this work.
 
 The plan of the paper is as follows. In section II we briefly sketch the set up
 of the photon emission rate calculation. In section IIIA the spectral representation of the effective fermion propagator is presented where we also show and discuss the dispersion curve.  The retarded photon self energy from chirally asymmetric medium is calculated in section IIIB. The imaginary part of the trace of the retarded photon self energy is evaluated in section IIIC. In section IV we present the final form of the photon production rate and show the numerical results followed by a brief summary in section V.

\section{Setup}
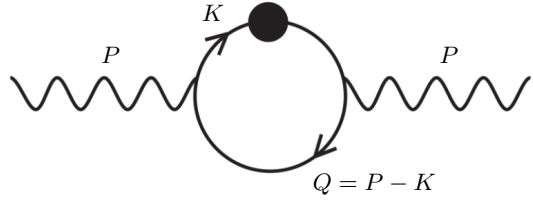
\begin{figure}
		\begin{overpic}[scale=0.26]{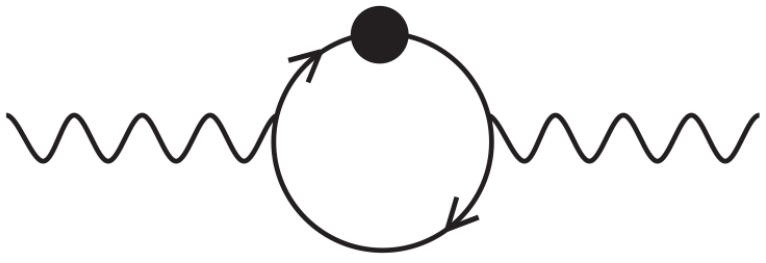}
			\put(58,-2){$Q=P-K$}
			\put(37,30){$K$}
			\put(18,22){$P$}
			\put(82,22){$P$}
		\end{overpic}
		\caption{The one-loop photon self-energy is evaluated in the HTL approximation. The internal fermion line with a black blob represents the soft quark propagator dressed with HTL resummation, while the line without the blob denotes the bare hard quark propagator. }
	\label{Fig_photon}
\end{figure}

The emission rate of real photons with energy $E$ and momentum $\vec{p}$ from a thermalized  plasma of quarks and gluons is expressed as~\cite{Kapusta:1991qp,Baier:1991em,Kapusta:2006pm,Bellac:2011kqa,Haque:2024gva}
\begin{equation}
	E \frac{dR}{d^3 p} = \frac{2}{(2\pi)^3} ~\IM~ ^{\rm ret}\Pi^\mu_\mu (E,\vec{p}) ~\frac{1}{e^{\beta E} -1} \label{Eq_Rate}
\end{equation}
where $^{\rm ret}\Pi^\mu_\mu$ is the retarded self-energy in a thermal medium. It is well known that this formula is exact to all orders in strong interaction and only up to leading order in electromagnetic interaction due to the assumption that the produced photon emerges without final state scattering from the matter i.e. the system size is small compared to mean free path of photons~\cite{Kapusta:1991qp,Kapusta:2006pm}. 
Here we only focus on the soft contribution to the photon rate as the hard contribution can be obtained employing perturbative methods. Since the energy of the produced photon is hard ($E \gg T$) there is only one dressed propagator  (represented by blob) which can be present kinematically in the polarization tensor since the other one has to be hard. As the photon carries a large momentum, it effectively probes the internal vertex structure eliminating the need for vertex corrections. By cutting this polarization tensor one effectively includes all processes through which the soft quark interacts with the medium. Thus we are led to evaluate the diagram shown in Fig.~\ref{Fig_photon} which is computed in the following sections.

\section{Photon self energy in chirally imbalanced medium}

Including $u$ and $d$ quarks the (trace of the) retarded photon self-energy as shown in Fig.~\ref{Fig_photon} can be  written as
\begin{equation}
	^{\rm ret}\Pi^\mu_\mu (P) = - \frac{5}{3}e^2 \SumInt_{\SB{K}} \Tr{\mS (K)\gm^\mu S(Q=P-K)\gm_\mu } \label{Eq_Self_Energy}
\end{equation}
with $\FSumInt{K} = T \sum_{k_0}\threeint{k}$ is a fermionic sum-integral with $k_0 = i \frac{(2n+1)\pi}{\beta}$ and the retarded self-energy implies $p_0$ has a small positive imaginary part. $\mS(K)$ is the effective quark propagator and $S(Q)$ is the bare quark propagator. Below we discuss the spectral representation of the effective fermion propagator in a chirally asymmetric medium.

\subsection{Spectral representation of quark propagator}
The effective quark propagator $\mS(K)$ in presence of chiral imbalance including the hard thermal loop in the quark self-energy is expressed as~\cite{Duari:2025iah}
\begin{equation}
	\mS (K) = \Pp \FB{\dfrac{\gm_0\Lambda_+}{ D_{L,+}(k_0^+,k)}+\dfrac{\gm_0\Lambda_-}{ D_{L,-}(k_0^+,k)}}\Pm + \Pm \FB{\dfrac{\gm_0\Lambda_+}{ D_{R,+} (k_0^-,k)}+\dfrac{\gm_0\Lambda_-}{D_{R,-}(k_0^-,k)}}\Pp \label{Eq_Full_Prop}
\end{equation}
where 
\begin{align}
	D_{L,\pm}(k_0, k ) &= k_0 \mp k  -\dfrac{M^2_L}{k} \TB{\dfrac{1}{2} \FB {1\mp \frac{k_0}{k}} \ln \dfrac{k_0 +k}{k_0-k} \pm 1}  \\
		D_{R,\pm}(k_0, k ) &= k_0 \mp k  -\dfrac{M^2_R}{k} \TB{\dfrac{1}{2} \FB {1\mp \frac{k_0}{k}} \ln \dfrac{k_0 +k}{k_0-k} \pm 1}  \\
		M_{L/R} &= M^2 \pm \delta M^2 = \dfrac{g^2 C_F}{8}\FB{T^2 + \frac{\mu^2}{\pi^2} + \frac{\mu_5^2}{\pi^2}} \pm  \dfrac{g^2 C_F}{4\pi^2} \mu \mu_5  ~.
\end{align}
The projectors $\Lambda_\pm = \half \FB{1 \mp \gm_0 \frac{\vec{\gm} \cdot \vec k}{k}}$ projects out spinors whose chirality is equal or opposite to the helicity and $k^\pm_0 = k_0+\mu\mp \mu_5$.  Here $M$ denotes the thermal mass of the quark while $\delta M $ represents an additional mass scale induced by the presence of chiral asymmetry. Since $\delta M$ is proportional to $\mu_5$ it is straightforward to verify that in absence of chiral imbalance $M_L = M_R = \dfrac{g^2 C_F}{8}\FB{T^2 + \frac{\mu^2}{\pi^2} }$ with $C_F=4/3$. It is convenient to introduce a spectral representation for the effective quark propagator. For any function of $k_0$ we have
\begin{equation}
	F(k_0) = \int_{-\infty}^\infty d \omega \dfrac{\rho (\omega)}{k_0 -\omega +i \epsilon}\label{Eq_def_SD}
\end{equation}
where $\rho$ is the spectral density function and can be determined by inverting the above relation which gives
\begin{equation}
	\rho(k_0) = -\frac{1}{\pi} ~\IM~F(k_0) = -\frac{1}{2\pi i} \text{ Disc}~F(k_0)~.
\end{equation} 
Using this one can obtain the spectral functions $\rho_{L,\pm}$ and $\rho_{R,\pm}$ corresponding to the dressed propagators $1/D_{L,\pm}$ and $1/D_{R,\pm}$ as
\begin{eqnarray}
	\rho_{L,\pm}(k_0^+,k)&= \dfrac{{k_0^+}^2-k^2}{2M^2_L}\TB{\delta\FB{k_0^+-\omega_{L,\pm}}+\delta\FB{k_0^++\omega_{L,\mp}}} + \beta_{L,\pm} (k_0^+,k) \theta \FB{k^2-{k_0^+}^2}~\label{Eq_rhoL}\\
	\rho_{R,\pm}(k_0^-,k)&= \dfrac{{k_0^-}^2-k^2}{2M^2_R}\TB{\delta\FB{k_0^--\omega_{R,\pm}}+\delta\FB{k_0^-+\omega_{R,\mp}}} + \beta_{R,\pm} (k_0^-,k) \theta \FB{k^2-{k_0^-}^2}~\label{Eq_rhoR}
\end{eqnarray}
where
\begin{eqnarray}
	\beta_{L,\pm} (k_0,k) &= \dfrac{\dfrac{1}{2 k} M^2_L \FB{1\mp k_0/k}}{\SB{k_0 \mp k  -\dfrac{M^2_L}{k} \TB{\dfrac{1}{2} \FB {1\mp \dfrac{k_0}{k}} \ln \MB{\dfrac{k_0 +k}{k_0-k}} \pm 1}}^2 + \dfrac{\pi^2 M^4_L}{4 p^2}\FB{1\mp \dfrac{k_0}{k} }^2}\\
	\beta_{R,\pm} (k_0,k) &= \dfrac{\dfrac{1}{2 k} M^2_R \FB{1\mp k_0/k}}{\SB{k_0 \mp k  -\dfrac{M^2_R}{k} \TB{\dfrac{1}{2} \FB {1\mp \dfrac{k_0}{k}} \ln \MB{\dfrac{k_0 +k}{k_0-k}} \pm 1}}^2 + \dfrac{\pi^2 M^4_R}{4 p^2}\FB{1\mp \dfrac{k_0}{k} }^2}
\end{eqnarray}
The $\delta$-functions in Eqs.~\eqref{Eq_rhoL} and \eqref{Eq_rhoR} come from the poles in $1/D_{L,\pm}$ and $1/D_{R,\pm}$ respectively and correspond to the propagation of $L$ and $R$-type quasiparticles~\cite{Duari:2025iah}. Concentrating on the $L$-mode, it can be seen that at positive energy $k_0 > 0$ the pole in the effective quark propagator has two branches. 
The first branch, $\omega_{L,+}(k)$, represents the propagation of ordinary $L$-quarks with a thermal mass $M_L$. This quasiparticle mode can be interpreted as a particle-like solution of $L$-modes as the ratio of its chirality to its helicity is $+1$. This is evident from Eq.~\eqref{Eq_Full_Prop}.  The second branch, $\omega_{L,-}(k)$ represents a collective mode which has no analogue at zero temperature. Since the ratio of its chirality and helicity is $-1$ this gives rise to a hole-like excitation, commonly referred to as the plasmino of $L$-mode.
An analogous interpretation holds about dispersion of the $R$-modes. 
\begin{figure}[h]
	\includegraphics[scale=0.275]{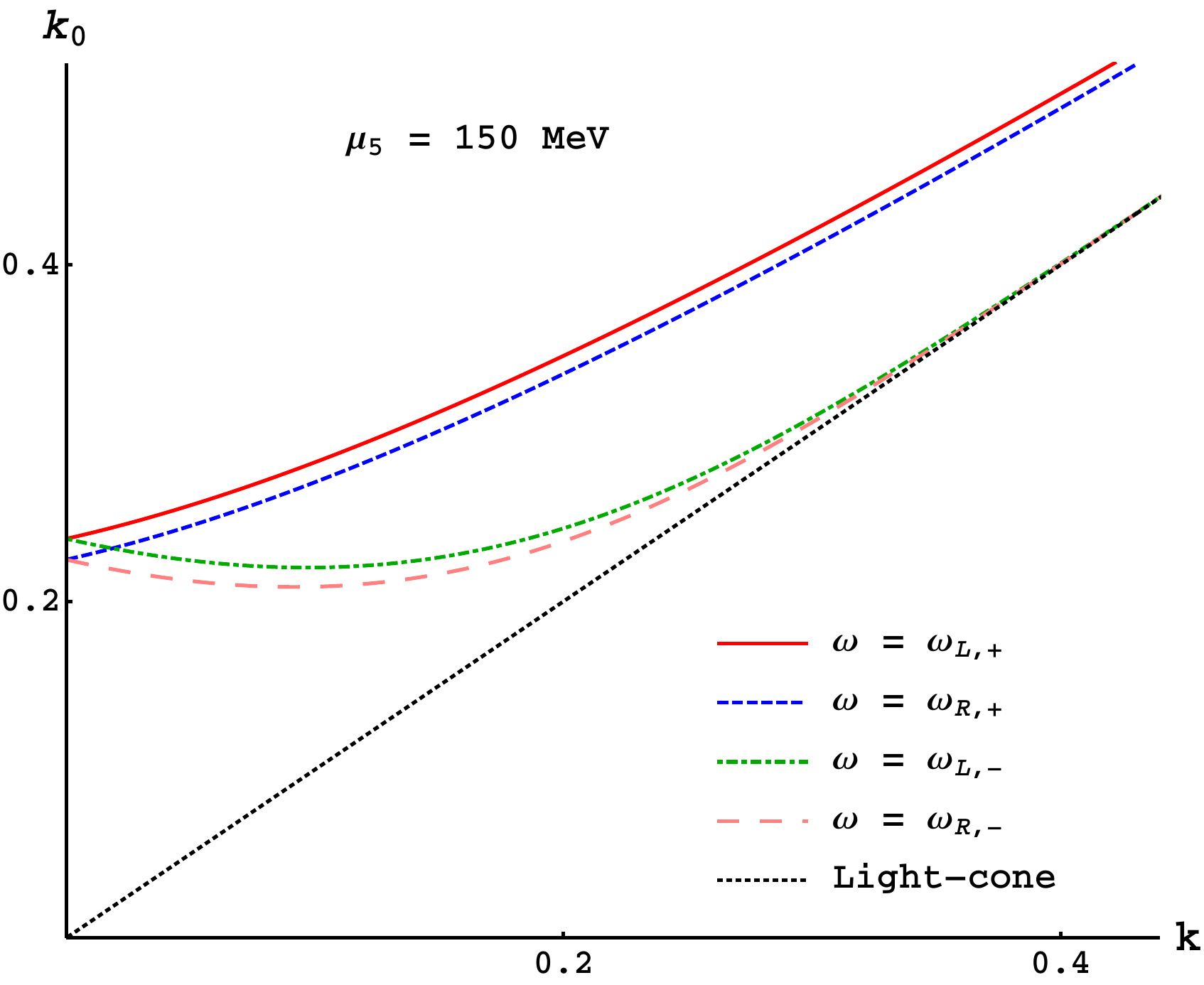}
	\caption{Dispersion of $L$ and $R$-mode for different values of $\mu_5$ at $T=200$ MeV and $\mu = 150$ MeV.}
	\label{Fig_Dispersion}
\end{figure}

Numerical solutions for a representative value of $\mu_5$ is shown in Fig.~\ref{Fig_Dispersion} illustrating the dispersion relations for both $L$ and $R$-modes. As can be seen, in the chiral plasma $L$ and $R$-modes disperse differently. At vanishing momentum the particle and plasmino solutions of $L$ and $R$-modes acquire different thermal masses denoted by $M_L$ and $M_R $ respectively. These are the fermionic analogues of the plasma frequency.  
At intermediate momenta, the plasmino branches of both $L$ and $R$-modes exhibit a characteristic non-monotonic dispersion. These modes can be interpreted as collective excitations arising from the mixing between positive-frequency fermion states and negative-frequency anti-fermion states induced by thermal effects. 
In particular, at small momentum the dispersion of the plasmino branches $\omega_{L/R,-}$  behaves in a similar way to a negative-energy mode as the energy decreases with increasing momentum.
A distinctive feature of the plasmino branch is the presence of a minimum at finite momentum. This non-monotonic behavior can be understood as follows. At $k =0$, the slope of the dispersion curve for the hole-like excitation is negative indicating that the group velocity is oppositely directed to the momentum. In contrast, for the particle-like excitation the group velocity has the same sign as the momentum. At large momentum, however, the plasmino branch asymptotically approaches the light cone  with the group velocity tending toward unity. The combination of these two limiting behaviors leads to a local minimum in the dispersion relation at finite momentum. The existence of such a minimum implies a vanishing group velocity at that point which can result in van Hove singularities in the density of states. These singularities in turn produce observable effects such as gaps or enhancements in physical quantities such as the dilepton production rate~\cite{Braaten:1990wp}.

In a similar way one can also express the bare propagator in spectral representation. It is convenient to first write the bare propagator in the same basis as the effective propagator as in Eq.~\eqref{Eq_Full_Prop}. Thus we obtain~\cite{Kharzeev:2009pj,Ghosh:2022xbf,Duari:2025iah}
\begin{eqnarray}
	S (Q) = \Pp \FB{\dfrac{\gm_0\Lambda_+}{ d_{L,+}(q_0^+,q)}+\dfrac{\gm_0\Lambda_-}{ d_{L,-}(q_0^+,q)}}\Pm + \Pm \FB{\dfrac{\gm_0\Lambda_+}{ d_{R,+} (q_0^-,q)}+\dfrac{\gm_0\Lambda_-}{d_{R,-}(q_0^-,q)}}\Pp \label{Eq_Bare_Prop}
\end{eqnarray}
where
\begin{align}
	d_{L,\pm}(q_0^+,q) = q_0^+\mp q ~~~~~~~~~~~~~~~\text{and}~~ ~~~~~~~~~~~~~ 	d_{R,\pm}(q_0^-,q) = q_0^-\mp q ~.
\end{align}
The spectral density functions for the bare propagators will be denoted by $\sigma$ and they are given by the following expressions
\begin{equation}
	\sigma_{L,\pm} = \delta \FB{q_0^+\mp q} ~~~~~~~~~~~~~~~\text{and}~~ ~~~~~~~~~~~~~\sigma_{R,\pm} = \delta \FB{q_0^-\mp q}~.
\end{equation}

\subsection{Calculation of in-medium photon polarization function at finite $\mu_5$}
Now using Eqs.~\eqref{Eq_Full_Prop} and \eqref{Eq_Bare_Prop} in Eq.~\eqref{Eq_Self_Energy} we get
\begin{align}
	\retPi &= - \dfrac{5}{3} e^2 \dfrac{1}{4}\SumInt_{\SB{K}}\TB{ \dfrac{\Tr{\gm_\mu\Pp \hat{\slashed{K}}\Pm \gm^\mu \Pp \hat{\slashed{Q}} \Pm }}{D_{L,+}(k_0^+,k)d_{L,+} (q_0^+,q)}+\dfrac{\Tr{\gm_\mu\Pp \hat{\slashed{K}}\Pm \gm^\mu \Pp \hat{\slashed{\ovQ}} \Pm }}{D_{L,+}(k_0^+,k)d_{L,-} (q_0^+,q)}
		+\dfrac{\Tr{\gm_\mu\Pp \hat{\slashed{\ovK}}\Pm \gm^\mu \Pp \hat{\slashed{Q}} \Pm }}{D_{L,-}(k_0^+,k)d_{L,+} (q_0^+,q)} \right.\nn \\ & \left.   \hspace{6 em} +
		\dfrac{\Tr{\gm_\mu\Pp \hat{\slashed{\ovK}}\Pm \gm^\mu \Pp \hat{\slashed{\ovQ}} \Pm }}{D_{L,-}(k_0^+,k)d_{L,-} (q_0^+,q)} + 
		\dfrac{\Tr{\gm_\mu\Pp \hat{\slashed{K}}\Pm \gm^\mu \Pp \hat{\slashed{Q}} \Pm }}{D_{R,+}(k_0^-,k)d_{R,+} (q_0^-,q)}
		+\dfrac{\Tr{\gm_\mu\Pp \hat{\slashed{K}}\Pm \gm^\mu \Pp \hat{\slashed{\ovQ}} \Pm }}{D_{R,+}(k_0^-,k)d_{R,-} (q_0^-,q)}  \right.\nn \\ & \left.\hspace{6em} +\dfrac{\Tr{\gm_\mu\Pp \hat{\slashed{\ovK}}\Pm \gm^\mu \Pp \hat{\slashed{Q}} \Pm }}{D_{R,-}(k_0^-,k)d_{R,+} (q_0^-,q)}  + 
		\dfrac{\Tr{\gm_\mu\Pp \hat{\slashed{\ovK}}\Pm \gm^\mu \Pp \hat{\slashed{\ovQ}} \Pm }}{D_{R,-}(k_0^-,k)d_{R,-} (q_0^-,q)}
	} \nn \\
	&= \dfrac{5}{3} e^2 \SumInt_{\SB{K}} \TB{
		\dfrac{1}{D_{L,+} (k_0^+,k)  }  \FB{\dfrac{\hat{K} \cdot \hat{Q}}{d_{L,+} (q_0^+,q) }  + \dfrac{\hat{K} \cdot \hat{\ovQ}}{d_{L,-} (q_0^+,q) }  }
		+ \dfrac{1}{D_{L,-} (k_0^+,k)  }  \FB{\dfrac{\hat{K} \cdot \hat{\ovQ}}{d_{L,+} (q_0^+,q) }  + \dfrac{\hat{K} \cdot \hat{Q}}{d_{L,-} (q_0^+,q) }  } 
		\right.\nn \\ & \left.\hspace{3em}
		+ \dfrac{1}{D_{R,+} (k_0^+,k)  }  \FB{\dfrac{\hat{K} \cdot \hat{Q}}{d_{R,+} (q_0^+,q) }  + \dfrac{\hat{K} \cdot \hat{\ovQ}}{d_{R,-} (q_0^+,q) }  }
		+ \dfrac{1}{D_{R,-} (k_0^+,k)  }  \FB{\dfrac{\hat{K} \cdot \hat{\ovQ}}{d_{R,+} (q_0^+,q) }  + \dfrac{\hat{K} \cdot \hat{Q}}{d_{R,-} (q_0^+,q) }  }
	}\label{Eq_Ret_Pi}
\end{align}
where $\hat{K}^\mu = \FB{1,\hat{k}},\hat{\ovK}^\mu = \FB{1,-\hat{k}}$ and we have used
\begin{align}
	\Tr{\gm_\mu\Pp \hat{\slashed{K}}\Pm \gm^\mu \Pp \hat{\slashed{Q}} \Pm } &= -4\hat{K} \cdot \hat{Q} = \Tr{\gm_\mu\Pp \hat{\slashed{\ovK}}\Pm \gm^\mu \Pp \hat{\slashed{\ovQ}} \Pm } \\
	\Tr{\gm_\mu\Pp \hat{\slashed{K}}\Pm \gm^\mu \Pp \hat{\slashed{\ovQ}} \Pm } &= -4\hat{K} \cdot \hat{\ovQ} = \Tr{\gm_\mu\Pp \hat{\slashed{\ovK}}\Pm \gm^\mu \Pp \hat{\slashed{Q}} \Pm } ~.
\end{align}
Note that in the limit $\mu_5 \to 0$, Eq.~\eqref{Eq_Ret_Pi} correctly reduces to the standard result of Ref.~\cite{Kapusta:1991qp} since the denominators for the $L$- and $R$-modes become identical.

\subsection{Imaginary part of the retarded photon self-energy}

We now evaluate the imaginary part of the retarded photon self-energy which is required for calculation of the photon production rate given by Eq.~\eqref{Eq_Rate}. Following Eq.~\eqref{Eq_Ret_Pi} we can write
\begin{align}
	\IM~ \retPi &= \dfrac{5}{3} e^2 \SumInt_{\SB{K}} \TB{
		\hat{K} \cdot \hat{Q} \FB{\IM~\dfrac{1}{D_{L,+} (k_0^+,k) d_{L,+} (q_0^+,q)} + \IM~\dfrac{1}{D_{L,-} (k_0^+,k) d_{L,-} (q_0^+,q)}+\IM~\dfrac{1}{D_{R,+} (k_0^+,k) d_{R,+} (q_0^+,q)} \right. \right.\nn \\&\left. \left. 
			\hspace{3 em} +  \IM~\dfrac{1}{D_{R,-} (k_0^+,k) d_{R,-} (q_0^+,q)}  }
		+\hat{K} \cdot \hat{\ovQ} \FB{\IM~\dfrac{1}{D_{L,+} (k_0^+,k) d_{L,-} (q_0^+,q)} + \IM~\dfrac{1}{D_{L,-} (k_0^+,k) d_{L,+} (q_0^+,q)} \right. \right.\nn \\&\left. \left. \hspace{3 em}
			+\IM~\dfrac{1}{D_{R,+} (k_0^+,k) d_{R,-} (q_0^+,q)}
			+ \IM~\dfrac{1}{D_{R,-} (k_0^+,k) d_{R,+} (q_0^+,q)}  }
	}~.\label{Eq_int_IMPI}
\end{align}
In the  above equation sum over Matsubara frequencies can be performed using the method discussed in~\cite{Braaten:1990wp} and for finite chemical potential the result is given by
\begin{equation}
	\IM~T\sum_{k_0} F_1(k_0) F_2(q_0 = p_0-k_0) = \pi \FB{e^{\beta p_0} -1} \int d \w d \wp  \rho_1(\w) \rho_2 (\wp) n^+_F (\w) n^-_F (\wp) \delta\FB{ p_0 -\w -\wp }~.
\end{equation}
Here $\rho_1$ and $\rho_2$ are the spectral density functions for $F_1$ and $F_2$ respectively and $n_F^\pm = 1/\FB{e^{\beta (\w \mp\mu) } +1}$ is the Fermi distribution function. Now using this result in Eq.~\eqref{Eq_int_IMPI} we get
\begin{align}
	&	\IM~\retPi = \dfrac{5 \pi}{3} e^2 \FB{e^{\beta p_0} - 1 } \threeint{k} \int d \w ~d \wp \nn \\
	& ~~~~~~\TB{
		n_F^{L,+} (\w)n_F^{L,-} (\wp)\SB{\hat{K} \cdot \hat{Q} \FB{\rho_{L,+} \sigma_{L,+}+\rho_{L,-} \sigma_{L,-} } +\hat{K} \cdot \hat{\ovQ} \FB{\rho_{L,+} \sigma_{L,-}+\rho_{L,-} \sigma_{L,+} }} \right. \nn \\ & \left.
		~~~~~~~+	~n_F^{R,+} (\w)n_F^{R,-} (\wp)\SB{\hat{K} \cdot \hat{Q} \FB{\rho_{R,+} \sigma_{R,+}+\rho_{R,-} \sigma_{R,-} } +\hat{K} \cdot \hat{\ovQ} \FB{\rho_{R,+} \sigma_{R,-}+\rho_{R,-} \sigma_{R,+} }}
	}~\delta\FB{ p_0 -\w -\wp }  \label{Eq_IMPI}~.
\end{align}
In the above expression, the spectral functions $\rho$ are evaluated at $(\omega, k)$, while the $\sigma$ functions are evaluated at $(\omega', q)$. Since we are interested in calculating the soft contribution, the exchanged quark propagator must be dressed and satisfies
\begin{equation}
	-k_c^2 \le \omega^2 - k^2 \le 0 \quad \implies \quad 0 \le k^2 - \omega^2 \le k_c^2 ,
\end{equation}
where $k_c^2$ denotes the cut-off in the four-momentum transfer $t$ (and $u$). This cut-off is introduced in the calculation of the hard contribution to regulate the infrared divergences arising from the soft region of phase-space. Consequently the delta-functions (corresponding to pole contributions) in the spectral densities $\rho_{L,\pm}$ and $\rho_{R,\pm}$ do not contribute at this order. Instead, only the branch-cut contributions represented by $\beta_{L,\pm}$ and $\beta_{R,\pm}$ are relevant. Thus we arrive at
\begin{align}
	&	\IM~\retPi = \dfrac{5 \pi}{3} e^2 \FB{e^{\beta p_0} - 1 } \threeint{k} \int d \w ~d \wp \nn \\
	& ~~~~~~~~~\TB{
		n_F^{L,+} (\w)n_F^{L,-} (\wp)\SB{\hat{K} \cdot \hat{Q} \FB{\beta_{L,+} \sigma_{L,+}+\beta_{L,-} \sigma_{L,-} } +\hat{K} \cdot \hat{\ovQ} \FB{\beta_{L,+} \sigma_{L,-}+\beta_{L,-} \sigma_{L,+} }} \right. \nn \\ & \left.
		~~~~~~~~~~~~+	~n_F^{R,+} (\w)n_F^{R,-} (\wp)\SB{\hat{K} \cdot \hat{Q} \FB{\beta_{R,+} \sigma_{R,+}+\beta_{R,-} \sigma_{R,-} } +\hat{K} \cdot \hat{\ovQ} \FB{\beta_{R,+} \sigma_{R,-}+\beta_{R,-} \sigma_{R,+} }}
	} ~\delta\FB{ p_0 -\w -\wp } \nn \\
	&=  \dfrac{5 \pi}{3} e^2 \FB{e^{\beta p_0} - 1 } \threeint{k} \int d \w   \TB{
		n_F^{L,+} (\w)n_F^{L,-} (p_0-\w)\SB{\hat{K} \cdot \hat{Q} \FB{\beta_{L,+} ~\delta\FB{ p_0 -\w -q }+\beta_{L,-} ~\delta\FB{ p_0 -\w +q }} \right. \right.  \nn \\ & \left. \left. 
			~~~~~~+ \hat{K} \cdot \hat{\ovQ} \FB{\beta_{L,+} ~\delta\FB{ p_0 -\w +q }+\beta_{L,-} ~\delta\FB{ p_0 -\w -q } } } +	~n_F^{R,+} (\w)n_F^{R,-} (p_0-\w)\SB{\hat{K} \cdot \hat{Q} \FB{\beta_{R,+} ~\delta\FB{ p_0 -\w -q }  \right.  \right. \right. \nn \\ & \left. \left.\left. ~~~~~~ + \beta_{R,-} ~\delta\FB{ p_0 -\w +q } }  +\hat{K} \cdot \hat{\ovQ} \FB{\beta_{R,+} ~\delta\FB{ p_0 -\w +q }+\beta_{R,-} ~\delta\FB{ p_0 -\w -q } }}
	}\label{Eq_IMPI1}~.
\end{align}

This can be simplified further by incorporating the following assumptions. Since the external photon is hard i.e. $p_0 \gg T$ and exchange quark is soft i.e. $\w \ll T$ we can write
\begin{equation}
	n_F^{L/R,+} (\w) \approx  \dfrac{1}{2} \quad \quad\quad  \text{and} \quad\quad \quad n_F^{L/R,-} (p_0 - \w) \approx  n_F^{L/R,-} (p_0 ) \approx e^{- \beta p_0} ~.
\end{equation}
Furthermore we note that $\delta$-function with argument $p_0 - \w + q$ will never go to zero due to kinematic reasons. The other $\delta$-function can be simplified using HTL approximation as
\begin{equation}
	\delta \FB{p_0 -\w -q}\approx \dfrac{1}{k} \delta \FB{ \cos \theta - \dfrac{\w}{k} } \quad \quad\quad  \text{since} \quad\quad\quad q= \MB{\vec{p}-\vec{k} } \approx p- k \cos \theta~.
\end{equation}
Using these  in Eq.~\eqref{Eq_IMPI1} we get
\begin{align}
	\IM~\retPi &= \dfrac{5 e^2}{12 \pi} \FB{e^{\beta p_0} - 1 }\dfrac{e^{-\beta p_0}}{2} \int_{0}^{k_c} dk \int_{-k}^{k} d\w \nn \\ & \hspace{5 em}~\TB{ (k-\w) \beta_{L,+} (\w,k)+(k+\w) \beta_{L,-} (\w,k) +(k-\w) \beta_{R,+} (\w,k)+(k+\w) \beta_{R,-} (\w,k) }~. \label{Eq_IMPI2}
\end{align}
The integrals involving $\beta_{L,\pm} $ and $\beta_{R,\pm} $ can be performed both numerically~\cite{Kapusta:1991qp} and analytically~\cite{Thoma:2000ne}. Thus finally we arrive at
\begin{align}
	\IM~\retPi &= \dfrac{5 e^2}{12 \pi} \FB{e^{\beta p_0} - 1 }\dfrac{e^{-\beta p_0}}{2} \TB{ M_L^2 \ln \dfrac{k_c^2}{2 M_L^2} + M_R^2 \ln \dfrac{k_c^2}{2 M_R^2}}\nn \\
	& \approx \dfrac{5 e^2}{12 \pi} \FB{e^{\beta p_0} - 1 } e^{-\beta p_0} M^2 \TB{  \ln \dfrac{k_c^2}{2 M^2} + \dfrac{\delta M^2}{ M^2} \ln \dfrac{M_R}{M_L} } \label{Eq_Final_IMPI}~.
\end{align}
\section{Soft contribution to the photon rate}

\begin{figure}[h]
	\includegraphics[scale=0.3]{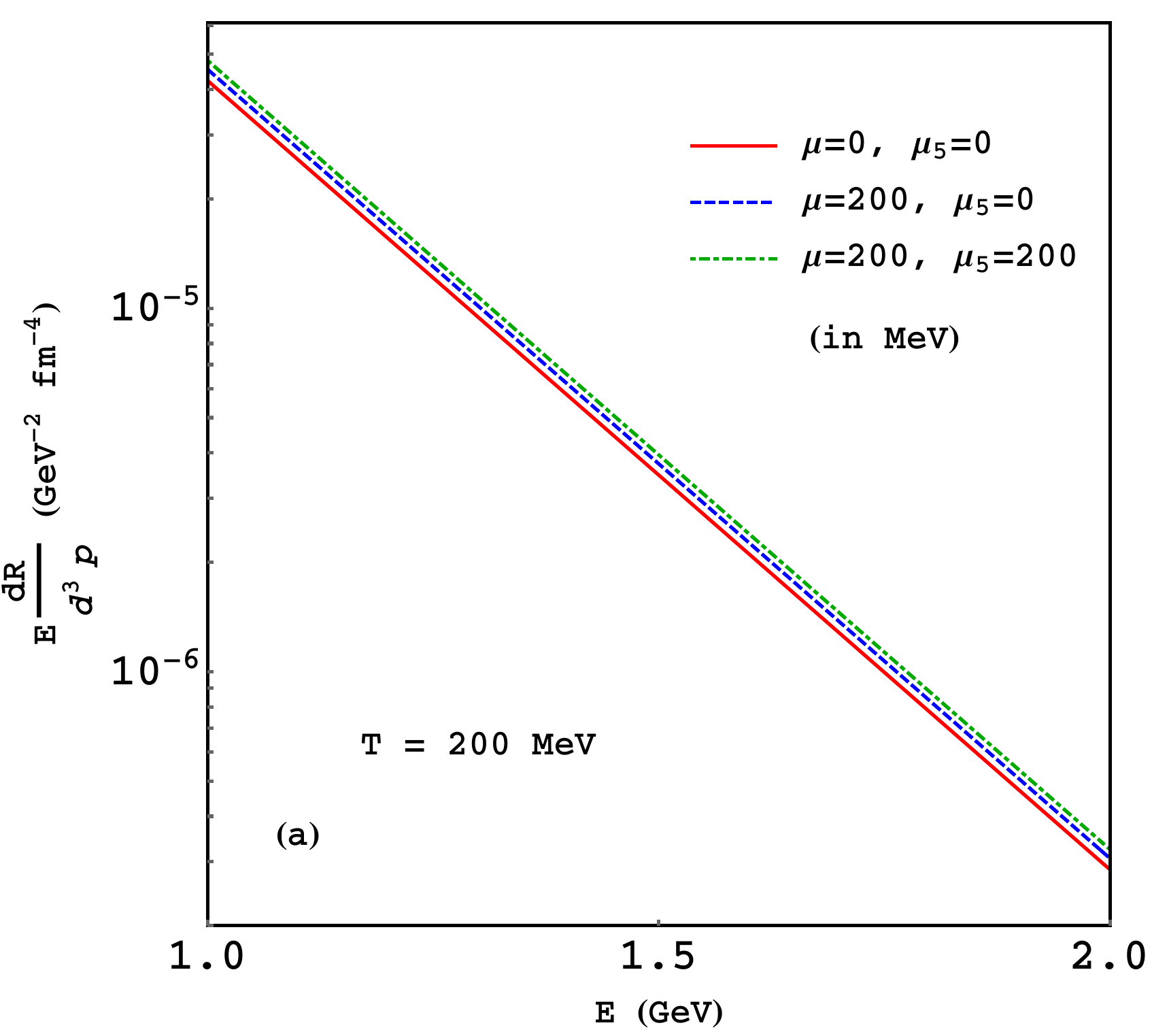}~~
	\includegraphics[scale=0.3]{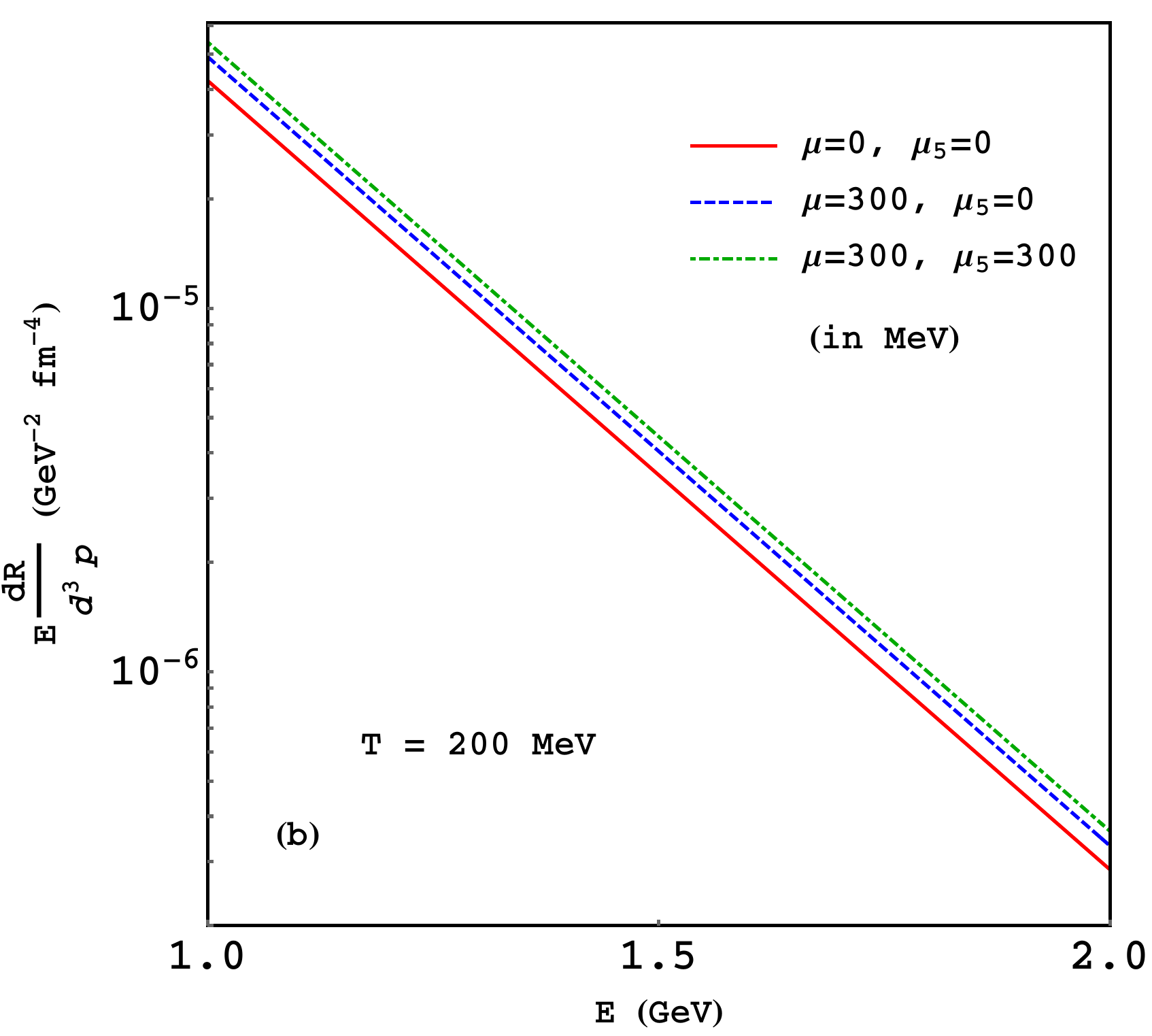}
	\caption{The photon production rate as a function of photon energy at $T$= 200 MeV and $\alpha_s = 0.3$. (a)  $\mu= 200$ MeV, $\mu_5 = 200$ MeV and  (b) $\mu= 300 $ MeV, $\mu_5= 300$ MeV
	}
	\label{Fig_rate}
\end{figure}

Now we can write down the final expression for the soft contribution to the thermal photon production rate from chirally imbalanced matter by using Eq.~\eqref{Eq_Final_IMPI} in Eq.~\eqref{Eq_Rate} 
\begin{equation}
	E\dfrac{dR}{d^3 p} = \dfrac{5}{9} \dfrac{\alpha \alpha_s}{2\pi^2} e^{-\beta E} \left(T^2+\frac{\mu^2}{\pi^2}+\frac{\mu_5^2}{\pi^2}\right) \TB{  \ln \dfrac{k_c^2}{2 M^2} + \dfrac{\delta M^2}{ M^2} \ln \dfrac{M_R}{M_L} }~. 
	\label{Phot_rate}
\end{equation}
Along with terms proportional to $\mu^2$ and $\mu_5^2$ the second term in the third bracket appear due to the presence of chiral imbalance.
In the limit $\mu_5 \to 0$, Eq.~\eqref{Phot_rate} correctly reduces to the result of~\cite{Traxler:1994hy} since $M_R, M_L \to M$. Furthermore at $\mu \to 0$ we recover the well known rate given in~\cite{Kapusta:1991qp}.  We plot in Fig.~\ref*{Fig_rate} the photon emission rates at $T=200$ MeV. Throughout this section we will choose $ \alpha_s = 0.3 $. The blue and green lines depict the enhancement due to the finite baryon chemical potential $\mu$ and chiral chemical potential $\mu_5$ respectively. By comparing the Figs.~\ref{Fig_rate} (a) and (b) we observe the dominant role played by $\mu$ which is in line with the observation in~\cite{Traxler:1994hy}. The presence of chiral imbalance represented by $\mu_5$ is seen to cause an additional enhancement in the rate.

A discussion on the choice of numerical values of $\alpha_s$ is in order here. At typical temperatures expected to be produced in QGP,   the value $\alpha_s = 0.3$ is consistent with a comparison between perturbation theory and lattice gauge simulations~\cite{Kapusta:1991qp,Kapusta:2006pm}. This implies $g\approx 2$ and our assumption of $g\ll 1$ while employing HTL approximation is strictly not valid. However as discussed in~\cite{Thoma:1993ii,Thoma:1993vs} the Braaaten-Pisarski method can be applied for large values of $g$ for the class of quantities which are logarithmically infrared divergent as in the present case, provided the energies or momenta of the particles are much greater than the temperature. Since we are working with production rate of hard photons with energies $ \sim$ GeV the extrapolation to realistic values of the coupling constant might be justified.

\begin{figure}[h]
		\includegraphics[scale=0.3]{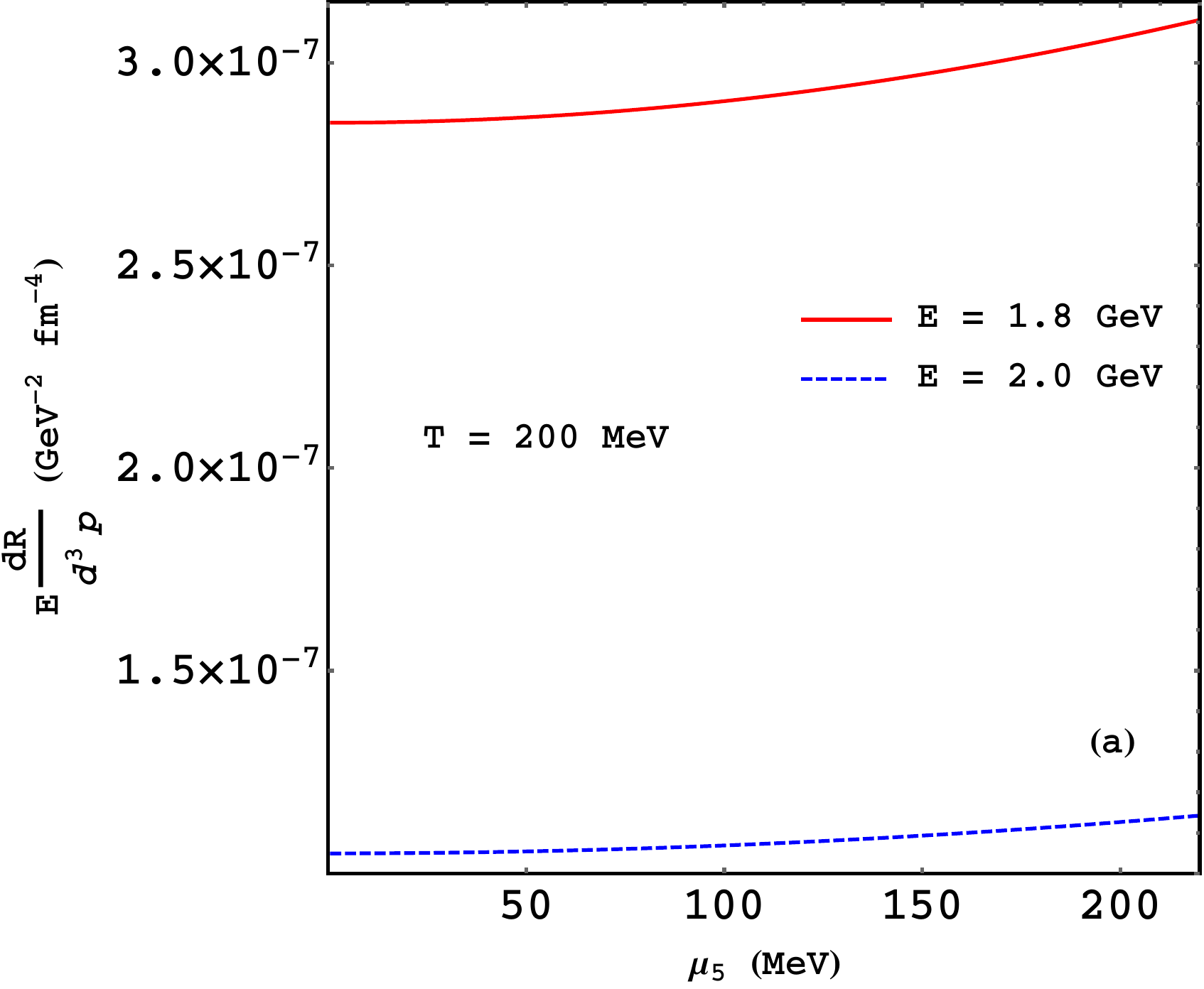}~~
		\includegraphics[scale=0.3]{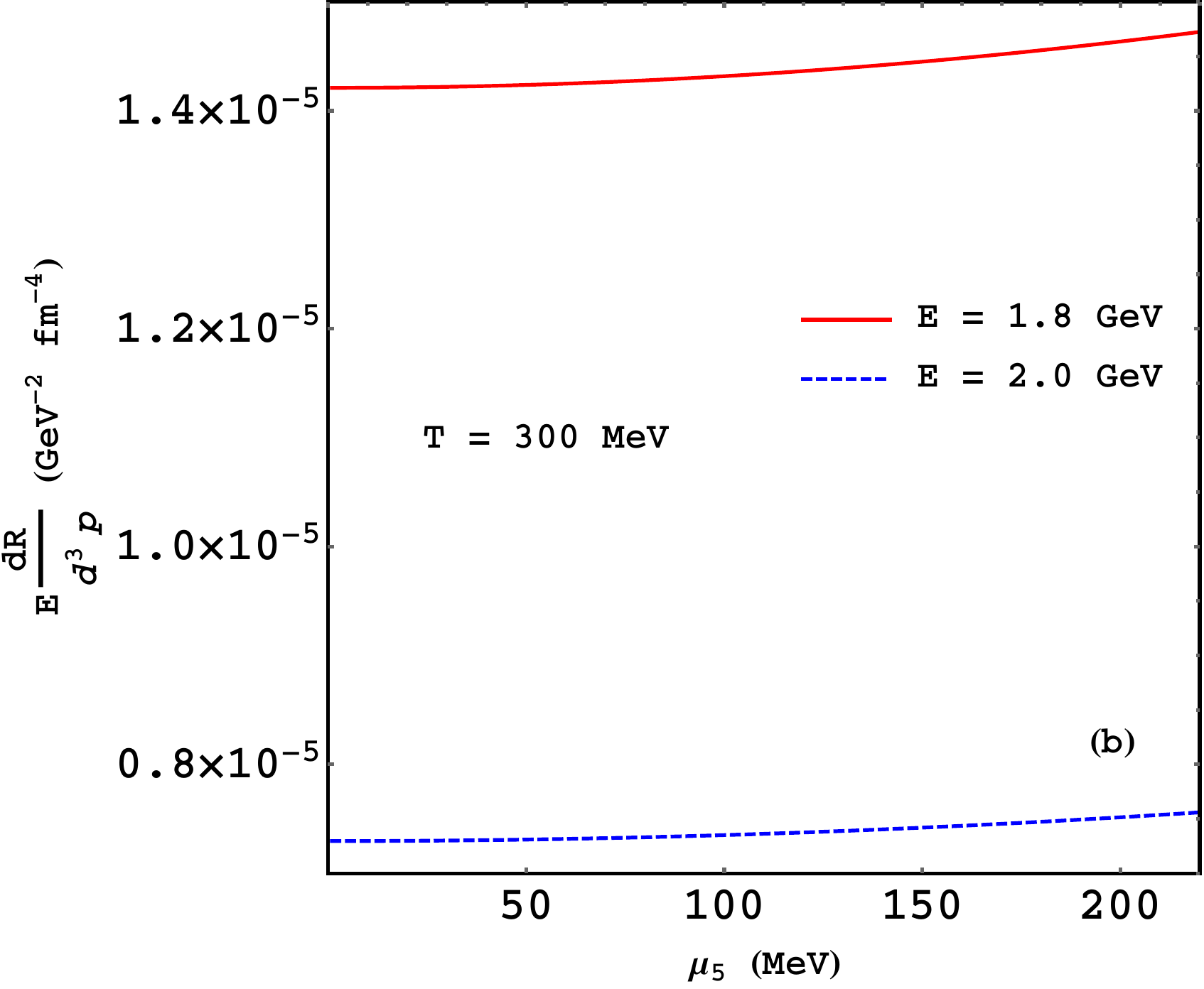}
		\caption{The photon production rate as a function of $\mu_5$ for different values of  photon energy at (a)  $T= 200$ MeV and  (b) $T= 300$ MeV in absence of $\mu$.}
		\label{Fig_rate_mu5}
	\end{figure}
We now focus specifically on how presence of chiral imbalance modifies the production rate of hard photons. For this we will put $\mu = 0$ and study the photon production rate as a function of $\mu_5$ for different values of temperature.  Such a scenario may be relevant for ultrarelativistic heavy ion collision experiments where the created matter  is expected to have baryon density close to zero but can have local $P$-odd domains with finite $\mu_5$.  In Fig.~\ref{Fig_rate_mu5} (a) we have shown the $\mu_5$-dependence of the production rate of hard photons at $T = 200 $ MeV considering energy of the photon $E = 1.8$ and 2 GeV. As evident from the plot, with increase in $\mu_5$ the rate increases and the enhancement is more prominent for lower values of energy. Qualitatively similar behaviour is observed in Fig.~\ref{Fig_rate_mu5} (b) where we have taken $T = 300$ MeV keeping all other parameters fixed.   Comparing Figs.~\ref{Fig_rate_mu5} (a) and (b) it can be inferred that with increase in temperature the production rate increases owing to the availability of a larger thermal phase space.

\section{Summary}

In this work we calculate the thermal photon emission rate from a quark gluon plasma with chiral imbalance. In particular, we have considered only the soft contribution to the photon self-energy by employing the dressed quark propagator obtained using the Hard Thermal Loop approximation. We first study
the quark dispersion relations and find that both the familiar quasiparticle and plasmino modes prevalent at finite temperature split into $L$ and $R$-modes in the presence of $\mu_5$ which disperse differently. At vanishing momentum the particle and plasmino solutions of the $L$ and $R$-modes acquire different thermal masses. 
At intermediate momenta, the plasmino branches of both $L$ and $R$-modes are found to exhibit a characteristic non-monotonic dispersion. Finally, the soft contribution to the thermal photon emission rate obtained from the retarded self-energy is found to contain additional terms proportional to the square of the quark and chiral chemical potentials which cause an enhancement to thermal photon emission in the presence of chiral imbalance. Though temperature is by far the leading scale, a notable enhancement in production rate with $\mu_5$ is observed for a given momentum. However in order to make an explicit connection to the experimentally observed photon excess a space-time evolution of the rate is to be performed using relativistic hydrodynamics with axial anomaly~\cite{Guo:2017jxs}. It is also to be noted that inclusion of a finite quark mass will lead to modification of the Dirac structure of the fermion self-energy and hence the dispersion relations, and may be implemented in chiral media following~\cite{Quimbay:1995jn}. Again, effects of momentum anisotropy as well as external magnetic field on photon production rate from a chiral plasma will indeed be interesting to study. However the complexity in the overall nature of the photon self-energy in this case due to new structure factors arising from the presence of additional vectors corresponding to the directions of the magnetic field and the momentum anisotropy is likely to render the calculation quite non-trivial.

\bibliography{reference} 

	\end{document}